\documentclass[floatfix,preprintnumbers,superscriptaddress,showpacs,
               showkeys,prb,preprint]{revtex4}

\usepackage{graphicx}
\usepackage{dcolumn}
\usepackage{amsmath, amsfonts, amssymb, bm}
\usepackage{psfrag} \usepackage{stdclsdv}

\begin{document}

\bibliographystyle{apsrev}

\title{
Hydrogen on Graphene Under Stress:
Molecular Dissociation and 
Gap Opening
}

\author{Hayley McKay}
\author{David J.~Wales}
\author{S.J. Jenkins}
\affiliation{
University Chemical Laboratories, Lensfield Road, Cambridge CB2 1EW, UK}

\author{J.A. Verges}
\author{P.L. de Andres}
\affiliation{
Instituto de Ciencia de Materiales de Madrid (CSIC)
E-28049 Cantoblanco, Madrid, SPAIN
}

\date{\today}

\begin{abstract}
Density functional calculations are employed to study the
molecular dissociation of hydrogen on graphene, the diffusion
of chemisorbed atomic species, and the electronic properties
of the resulting hydrogen on graphene system. 
Our results show that applying stress to the graphene substrate
can lower the barrier to dissociation of molecular hydrogen
by a factor of six, and change the process from endothermic to exothermic.
These values for the barrier and the heat of reaction, unlike
the zero stress values, are compatible with the time scales observed
in experiments. 
Diffusion, on the other hand, is not greatly modified by stress.
We analyse the electronic structure for
configurations relevant to molecular dissociation and
adsorption of atomic hydrogen on a graphene single layer.
An absolute band gap of $0.5$ eV is found for the equilibrium
optimum configuration for a narrow range of coverages 
($\theta \approx 0.25$). This value is in good agreement
with experiment [Elias et al., Science {\bf 323}, 610 (2009)].
\end{abstract}



\maketitle

\section{Introduction}
The safe and efficient storage of hydrogen is a crucial step
towards its use in the future as an energy vector.
The U.S.~Department of Energy (DOE) has launched an important
challenge by funding research to overcome the threshold of
6.5\,kg of stored hydrogen per 100\,kg of total storage system
weight.
\cite{patchkovskii05}
In the present contribution we focus on the
`gas adsorbed on solid' approach, with graphene layers
as the supporting material.
Adsorption of hydrogen on graphene (G) can take place in different
ways: (i) physisorption of molecular hydrogen, (ii) physisorption
of atomic hydrogen, (iii) chemisorption of atomic
hydrogen, (iv) chemisorption of small clusters of atomic hydrogen.
These scenarios exhibit different adsorption/desorption energies, optimum
geometrical configurations and diffusion barriers.
A good understanding of the interplay
between molecular physisorption and atomic chemisorption
would be useful for addressing the problem of hydrogen storage, as
well as for understanding fundamental issues, such as the puzzle of
how H$_{2}$ is formed in the interstellar medium.\cite{hornekaer06a}

Graphene itself is currently of particular interest
due to the recently developed capability for obtaining
samples consisting of
single or a few layers.\cite{novoselov04}
Such layers provide an excellent model
for advancing the applications of related materials,
such as graphite and nanotubes.
In particular, graphite
is an attractive candidate for hydrogen storage
because it is cheap, chemically inert and environmental friendly.
However, theoretical and experimental studies have
cast doubts about achieving
the goal proposed by the DOE, with certain exceptions
that have focused in the quantum properties of such a light element as
hydrogen.\cite{patchkovskii05}

To investigate the feasibility of loading and releasing hydrogen
efficiently in graphite or carbonaceous minerals we have characterised
diffusion and dissociation barriers for hydrogen on graphene
under stress.
External and internal stresses have been reported to induce interesting 
chemical properties in
carbon-based molecules,\cite{hoffmann08}  nanotubes,\cite{ruoff99}
graphene layers,\cite{deandres08b} etc.
We find that on applying a moderate amount of stress the
equilibrium state and the barriers for molecular dissociation
become much more favourable. 

Understanding the hydrogenation of graphene layers
is important from several perspectives. Recently,
in a series of elegant experiments, Elias et al.\cite{elias09} have
demonstrated the reversible transition of graphene layers from
a semimetallic to a semiconducting regime upon adsorption of hydrogen.
This is an interesting observation, which should help to advance the
application of graphene layers in microelectronics.
Our results should provide insight into these experiments, particularly
the conditions required to create an absolute band gap in the graphene
semi-metallic band structure, in terms of the adsorbate coverage.

\subsection{Methods}

Our calculations are based on
density functional theory (DFT).\cite{hohenberg64}
We calculated the adsorption energies, dissociation pathways, and barriers
for atomic and molecular hydrogen on
a $3 \times 3$ periodic supercell
with a vacuum gap in the perpendicular direction of $10$ to $20$ {\AA}.
The wavefunctions were expanded in a plane-wave basis set
up to a cutoff of $350$\,eV and were sampled on
a Monkhorst-Pack $6 \times 6 \times 1$ mesh inside the Brillouin zone.
Electronic bands were obtained using a smearing width of
$\eta=0.01$\,eV.
Carbon and hydrogen atoms were described by soft
pseudopotentials.\cite{vanderbilt90}
The choice of the exchange and correlation (XC) potential is
an important aspect of DFT calculations.
We considered both
the local density approximation (LDA)\cite{kohn65} and
a generalized gradient approximation (GGA) functional, RPBE,\cite{RPBE}
to check the influence of that choice on our results.
This formalism is accurate enough to reproduce structural and electronic
properties of chemisorbed species, but cannot accurately describe the
physisorption regime, because the
non-local correlation effects needed to properly describe weak van der Waals
interactions are not included.
The energy differences of interest here,
and their variation with external stress, are relatively insensitive
to whether the LDA or RPBE functional is used.
Furthermore, the errors associated with
the DFT formalism at large distances are unimportant, because the energies
involved in making/breaking chemical bonds are much greater than the few
meV associated with van der Waals interactions.

Total energies and gradients were computed with the CASTEP
program,\cite{payne05} interfaced to OPTIM\cite{optim}
for geometry optimisation.
All the transition states reported below were
refined using the gradient-only version of hybrid
eigenvector-following\cite{MunroW99,KumedaMW01,Wales03}
with a convergence condition of $0.01$\,eV/{\AA}
for the root mean square gradient.
Approximate steepest-descent paths were calculated by energy minimisation
for each transition state to characterise the corresponding pathway using
the limited-memory Broyden--Fletcher--Goldfarb--Shanno (LBFGS)
algorithm,\cite{liun89}
following displacements of order $0.01$\,\AA\
parallel and antiparallel to the Hessian eigenvector corresponding to the
unique negative Hessian eigenvalue.

\section{Results and Discussion}

The most favourable configuration for hydrogen on graphene corresponds to
molecular physisorption $3.5$\,{\AA} above the hollow site.
The physisorption energy is small,\cite{bonfanti07}
but it is not relevant here, since we are
dealing with processes involving much larger changes in
energy, such as the formation of a hydrogen-carbon bond,
or breaking the bond in the hydrogen molecule.
In contrast, for atomic hydrogen the most favourable configuration
involves chemisorption directly on top of a carbon atom.
RPBE calculations show that
the C-H bond length is then $1.1$\,{\AA}, corresponding to an adsorption
energy of 
$E_{G+H}-E_{G}-\frac{1}{2}E_{H_{2}}=+1.68$\,eV, while the carbon
directly below is puckered upwards by $0.4$\,{\AA}.
It is important to notice that
chemisorption of atomic hydrogen interferes
with the strong sp$^{2}$ sigma bonds
of planar graphene, producing an
energetic cost derived from the elastic deformation of the substrate.
Hydrogen atoms approaching a graphene surface from the gas phase
therefore encounter a barrier in reaching the chemisorbed state.
This barrier depends on the position on the surface,
varying between $0.28$\,eV and $1.27$\,eV
for atop and hollow sites (RPBE functional).

\begin{figure}[ht]
\includegraphics[width=0.99\columnwidth]{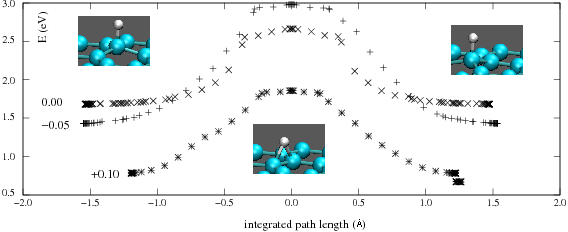}
\caption{
Energy/eV (RPBE functional\cite{RPBE}) 
as a function of integrated path length (in {\AA}) for
diffusion of chemisorbed atomic hydrogen on graphene
under different uniform stresses characterised by
the corresponding unit cell strains
($\epsilon=+0.1$, tensile; $\epsilon=-0.05$,
compressive). The structures of the transition state and minima are
superimposed.
}
\label{FigDiff}
\end{figure}

Both atomic and molecular hydrogen are trapped on the graphene
surface in a relatively shallow physisorption well at large distances
($\ge 3.5$\,{\AA} for the RPBE functional).
The diffusion barriers for physisorption are low, and hence the adsorbed
species
are mobile.
However, physisorbed species are subject to a correspondingly high
desorption probability;
at room temperature (RT) physisorbed atomic or molecular hydrogen desorb
on a time scale of nanoseconds.
The only way to keep such a system stable on a reasonable time scale
for practical applications is to reach a chemisorbed state.
While hydrogen molecules do not chemisorb, atomic hydrogen can form
strong bonds on atop sites, provided that
(i) the molecule initially dissociates and (ii)
atoms can find the favourable
region where barriers between physisorbed and
chemisorbed minima are small.
The values we obtain for the barriers imply that at RT a
hydrogen atom would find its way to the chemisorbed well on a time scale of
nanoseconds
near the atop region,
while it would take years for chemisorption to occur via a trajectory
only involving the local minimum corresponding to the hollow site.
To meet the DOE's challenge we need to consider the
formation and dissociation of the hydrogen molecule in the high
coverage regime.
The reaction H$_{2}$ $\rightarrow$ 2\,H near a free-standing graphene
sheet is associated with a barrier of $3.29$\,eV and an endothermic
heat of reaction of $1.89$\,eV,
making dissociation of molecular hydrogen a rare event
(RPBE values quoted here).
Furthermore, to load a stack of graphene layers efficiently
we need to consider  the diffusion barriers for atomic hydrogen.
Our calculations show that chemisorbed hydrogen diffuses
from top site to top site via bridge sites with an associated barrier of $0.98$\,eV.
External stress increases the diffusion barrier by $9$\% and $60$\% for
tensile and compressive strains of $\epsilon=0.1$ and $-0.05$, respectively (Figure \ref{FigDiff}).
These barriers result in diffusion times that are too long for effective
loading of chemisorbed hydrogen in graphite at RT
($\approx (10^{4}$\,s for a single hop),
leaving diffusion of molecular
hydrogen or physisorbed atomic species as the most practical scenarios.
Diffusion through the hollow site is unfavourable, 
involving a barrier of $4.05$\,eV,
resulting in a practically impenetrable layer in the perpendicular
direction at RT, in good agreement with experiment.\cite{elias09}

\subsection{H$_{2}$ Dissociation on Graphene}

\begin{figure}[ht]
\includegraphics[width=0.99\columnwidth]{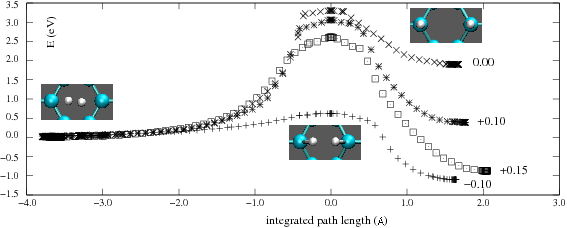}
\caption{
Energy profile/eV  (RPBE functional\cite{RPBE}) 
as a function of integrated path length (in {\AA})
for H$_{2}$ $\rightarrow$ $2$H under different
compressive/tensile strains, $\epsilon$.
The structures of the transition state and minima are superimposed.
}
\label{FigBarriers}
\end{figure}

We now examine the effect of internal and external stress on
H$_{2}$ dissociation/formation.
Figure \ref{FigBarriers} shows how the barriers and reaction
energies for the dissociation process are affected by
external stress.
The deformations considered here are quite realistic
for graphene, a material that can accommodate 
tensile strains up to $\epsilon=0.25$  
(corresponding to a stress of $\sigma \approx 42$ Nm$^{-1}$) before breaking up.\cite{lee08}
Compressive strains of up to $\epsilon=-0.05$ have also been 
measured from diffraction experiments in hydrogenated 
samples\cite{elias09}
(our calculations corroborate that these samples tend to buckle on an
atomic scale, at least in the region around the adsorbates).
Indeed, C-C bonds can accommodate a wide range of values,
from $1.37$\,{\AA} for triple bonds (sp hybridisation) to $1.54$\,{\AA} for
single bonds (sp$^{3}$ hybridisation), with a typical minimum value
in allenes of $1.30$\,{\AA} and a maximum
observed bond length of $1.78$\,{\AA}.\cite{hoffmann08}
For tensile stresses related to uniform deformations in the 2D lattice
between $\epsilon=0.10$ and $0.15$ the barriers are
not strongly affected, but the heat of reaction for molecular
hydrogen dissociation changes from
endothermic ($\Delta E= 1.89$\,eV for $\epsilon=0$) to exothermic
($\Delta E = -0.88$\,eV at $\epsilon=0.15$).
These results (values quoted for RPBE) show that between RT 
and $500$\,K, physisorbed
molecular hydrogen would not transform to chemisorbed atomic hydrogen on
a reasonable time scale.
In contrast, a compressive strain introduces a totally different scenario:
for $\epsilon=-0.05$ the RPBE barrier drops to $2.19$\,eV, improving the 
predicted time for dissociation by about a factor of $10^{11}$; 
for $\epsilon=-0.1$ 
the barrier decreases to $0.59$\,eV,
which should allow a single physisorbed H$_{2}$ molecule to
dissociate on a time scale of milliseconds at RT.
These RPBE values are expected to compare well with experiment, since
good accuracy has been obtained for similar systems,
with a slight tendency to overestimate the barriers.\cite{RPBE}
To assess the influence of the XC model we compare with LDA,
to obtain lower bounds for these barriers.
The LDA barrier for $\epsilon=0$ ($2.38$\,eV) drops 
to $1.4$\,eV under a compressive strain of $\epsilon=-0.05$,
and to $0.04$\,eV under a compressive strain of $\epsilon=-0.1$.
The rate of change of the barrier with respect to external stress
is therefore similar for both functionals (changes of $2.70$\,eV and $2.34$\,eV for
$\Delta \epsilon = -0.1$),
highlighting the consistency of predicted trends derived from 
introducing stresses to the system.
Different XC models produce slightly different
values for the strain threshold necessary for dissociation to 
occur on a given time scale, but 
our best estimate of $\epsilon=-0.05$ to $-0.1$ is narrow enough
to be significant.
Furthermore, we remark that the reaction would quickly reach equilibrium, due
to the more favourable exothermic heat of reaction for compression
(Figure \ref{FigBarriers}, lower curve).

We therefore predict that the pathway for molecular dissociation near stressed
graphene enables the reaction to occur much faster.
Elias {\it et al.}~have recently demonstrated that a single graphene layer
can be hydrogenated to open a band gap of about $0.5$\,eV.\cite{elias09}
Transmission electron microscopy provides evidence for a non-uniform
distribution of stresses, resulting in measurable variations of the 2D unit
cell parameter ranging from $-0.05$\% (compressive) to $0.03$\% (tensile).
Hydrogenation is performed by exposing graphene to a low pressure
hydrogen-argon mixture (10\% H$_{2}$) at about $600$\,K.
This process is reversible, and mass spectrometry has been used to
detect molecular hydrogen leaving the substrate after
annealing to about the same temperature.
From our results we interpret the experimental
evidence in the following way.
Starting with physisorbed molecular hydrogen,
a few molecules dissociate to produce chemisorbed
atomic hydrogen on a time scale of
$\tau \approx 1/(10^{16}\times10^{12}\times e^{(-B/k_ {B}T)})$,
where $10^{12}$\,Hz represents a typical attempt frequency,
$10^{16}$ is the number of molecules in contact with the surface
per mole of hydrogen,
and the Boltzmann factor defines the dependence on the
barrier, $B$, for an effective surface temperature, $T$.
At about $300^\circ\,$C a few molecules should start to dissociate on
a time scale of seconds, and the chemisorbed hydrogen induces
buckling on the surface, creating a local region where
the substrate is under an effective compressive stress.
According to our results, this stress lowers the barrier for dissociation
(Figure \ref{FigBarriers})
and helps to dissociate other molecules,
the chemisorbed atomic hydrogen cannot diffuse away 
at a noticeable rate
and a cluster of atomic hydrogen is formed around that region.
Atomic hydrogen is therefore preferentially adsorbed in strained regions.
Strains and stresses derived from the adsorption of hydrogen
are well inside the elastic regime, and graphene layers can
recover their original geometry after annealing to $300^\circ\,$C,
when atomic hydrogen desorbing from the surface recombines to
form molecular hydrogen, which is the species detected.
This picture is supported by STM experiments, where
clustering of hydrogen atoms due to preferential
sticking has been observed.\cite{hornekaer06b}
In this interpretation, the mechanism responsible for the
preferential sticking is the modification of C-C
bonds under local stress induced by neighbouring
chemisorption of hydrogen.
The scenario proposed here is physically reasonable and backed up by
state-of-the-art calculations that indicate how dissociation
can occur on a time scale of days rather than years. 

It has been suggested that clustering is related to
the elastic energy penalty paid for clustered hydrogen vs.~maximally dispersed 
hydrogen at sub-monolayer coverages.
A careful consideration of this point is beyond the scope of this work,
but nevertheless we can estimate the effect of clustering on
the elastic deformation of the substrate by comparing total energies
of the carbon atoms alone frozen in their equilibrium positions 
due to the presence of hydrogen. 
In particular, in a $2 \times 2$ supercell,
chemisorbing a single hydrogen produces an elastic deformation in
the substrate that
requires $0.88$\,eV. The corresponding deformation for two
hydrogen atoms located in their optimum position (nnNN site) 
requires $2.09$ eV. 
Therefore, positioning a second hydrogen in
the vicinity of the first one 
carries a significant energetic penalty,
and suggests to us that clustering is better
explained in terms or our proposed mechanism
than by a balance of elastic deformations. 
However, further work will be required to resolve this question.

\subsection{Electronic Structure Effects}

\begin{figure}
\includegraphics[width=0.95\columnwidth]{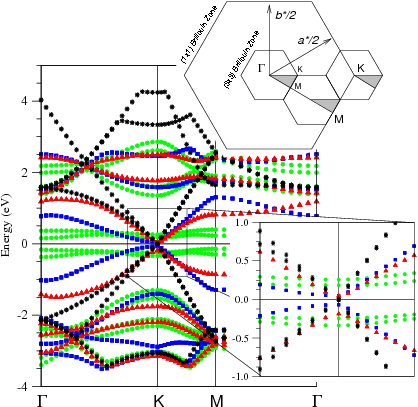}

\caption{
Electronic band structure for clean graphene and
hydrogen adsorbed on graphene ($\theta=1/9$):
Clean graphene (circles, black),
H+H (nnNN)/graphene (triangles-up, red),
H+H(nNN)/graphene (squares, green),
H+H (NN)/graphene (triangles-down, blue).
The central zone around {\bf K} has been magnified
on the right.
Inset: reciprocal-space path ($\Gamma$-{\bf K}-{\bf M}-
$\Gamma$) in the $3 \times 3$
irreducible Brillouin zone
and its relationship to the $1 \times 1$ zone.
}
\label{FigBS}
\end{figure}

\begin{figure}
\includegraphics[width=0.99\columnwidth]{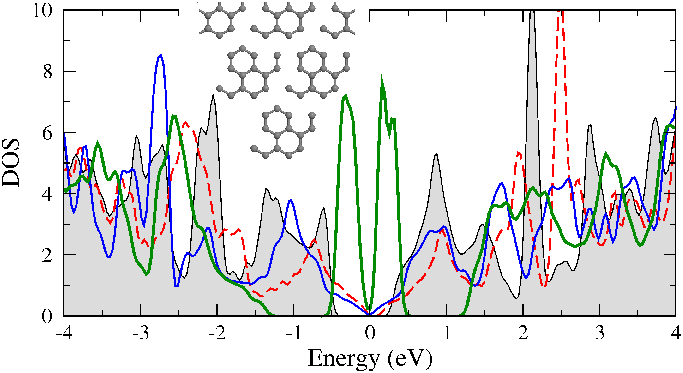}
\caption{
Density of states (eV$^{-1}$, smearing width $\eta=0.05$\,eV )
for the following configurations
(coverage, $\theta=1/9$):
nnNN (red, short-dashed), nNN (green, dotted) and
NN (blue, dot-dashed).
The shaded curve corresponds to nnNN at
$\theta=2/9$ where an
absolute gap of $0.63$\,eV appears, and
the inset illustrates the connectivity
for this case.
The Fermi energy is aligned at the origin.
}
\label{FigDOS}
\end{figure}

After molecular dissociation and chemisorption of the individual 
hydrogen atoms
the characteristic graphene semi-metallic electronic band structure 
is perturbed and partial gaps open 
in some directions in the Brillouin zone.
However, for the system to acquire semi-conducting properties, an absolute
gap must appear in the density of states (DOS).
We find that such an absolute gap is sensitive to the coverage,
which allows us to provide an estimate for comparison with experiment.
The origin of this gap is related to
the periodic potential created by the accumulation of charge
around the C-H bond and/or by the electronic band shifts originating
in the local modifications of the C-H bonding orbital.\cite{duplock04}
Figures \ref{FigBS} and \ref{FigDOS} show the
band structure and density of states corresponding to the different
configurations we have considered for two chemisorbed hydrogen atoms
in a $3 \times 3$ cell ($\theta = 1/9$).
Here we describe the LDA results, but the GGA calculations are essentially the same.

For clean graphene near the {\bf K} point we observe the distinctive
linear crossing of bands at the Fermi energy responsible
for semi-metal character.
Physisorbed atomic and molecular hydrogen
do not significantly change this behaviour.
The next interesting feature is the appearance of a half-filled
non-dispersing band at the Fermi level upon chemisorption of
a single hydrogen atom.\cite{duplock04}
Adsorption of a second hydrogen atom after molecular dissociation
can occur in three neighbouring sites:
(i) nearest-neighbour position (NN),
(ii) next-to-NN position (nNN),
or (iii) next-to-next-NN (nnNN).
According to Lieb's theorem, for a bipartite lattice like graphene
the ground state for NN, nNN and nnNN configurations is attained
with $S = 0, 2$, and 0\,$\mu_{B}$, respectively.\cite{lieb89}
Although Lieb's theorem is
only strictly valid for the Hubbard Hamiltonian,
we have found that it correctly predicts
the optimum spin configurations obtained
using DFT.
The global minimum is related to
occupancy of nnNN sites, while the next most stable is
the NN position. The intermediate site (nNN) is the
least stable one; it is a magnetic solution,
and spontaneously evolves to nnNN under the
restriction of $S = 0$. From an electronic point of view,
bands associated with the nNN site appear similar to the ones
for only one hydrogen atom, showing that
direct interaction between atoms at this distance is weak.
However, interaction between adsorbates mediated by the substrate
through static elastic distortions can be
important and should not be neglected.
These distortions produce a departure from a flat
geometry and are favourable, so long as they promote
the sp$^{2}$ to sp$^{3}$ transformation.
For high coverages, chemisorption on nnNN sites opens an absolute gap
allowing the system to find a more stable configuration.
The absolute band gap evolves from $3.1$\,eV for
$\theta=1/4$ to $0.6$\,eV
for $\theta=2/9$.
Below and above these two values for the hydrogen coverage
the computed band gap is too small or too large 
compared with experiment. Therefore,
we conclude that the actual coverage in ref.~\onlinecite{elias09}
should be near $\theta \approx 1/4$, and the geometrical
configuration near to the optimum equilibrium one.
It is interesting to notice that
occupation of the alternative locations (NN and nNN) is
not only a non-optimal configuration, but more crucially it does not
result in an absolute gap in the density of states even for the
larger coverages considered ($\theta=1/4$).
A characteristic double peak appears (Figure \ref{FigDOS}),
similar to that found for chemisorption of a single hydrogen atom,
where only a direct gap at the {\bf K} point
appears (not an absolute one).\cite{duplock04}
These results can be understood in terms of topological connectivity:
if we assume that the C-H bond saturates the p$_{z}$ electron
on that site,
occupation of nnNN sites for $\theta \ge 2/9$  breaks the
graphene layer in a set of nearly disconnected hexagons where
first-neighbour interactions are too weak and result in a series of
narrow peaks and an absolute gap
(inset and shaded curve in Figure \ref{FigDOS}).
For $\theta=1/4$ the rings become totally disconnected and
a larger gap appears.
This kind of reconstruction is reminiscent of a Peierls distortion,
where the system minimises its energy by a rehybridization of the
basis in the unit cell under a new balance between kinetic and
potential energy contributions.
While the popular
$2 \times 2$ tight-binding Hamiltonian including only $\pi$ electrons
reproduces most of the low energy physics correctly, a proper account
of the above effect requires consideration of the in-plane sp$^{2}$ 
sigma bonds
responsible for the elastic contribution to the energy.

\subsection{Conclusions}

In summary, we have studied the effect of external stress on diffusion of
atomic chemisorbed hydrogen and on the dissociation
of the H$_ {2}$ molecule. While external
stress has little influence on diffusion barriers we find that
compression changes the endothermic dissociation
of physisorbed H$_ {2}$ to exothermic, and lowers the barrier
by about $2.5$\,eV ($\epsilon=-0.1$)
for both the LDA and GGA functionals
considered.
Dissociation of H$_2$ is the first step for activated
chemisorption of hydrogen atoms, and leads to a significant
strain in the substrate. This strain lowers the barrier for
further chemisorption, favouring the pairing of chemisorbed species over
a small region related to the elastic distortion created by
the formation of the first C-H bond.
The most favourable configuration for a pair of chemisorbed
hydrogens is a paramagnetic ground state occupying
the next-to-next-nearest-neighbour site, while
the second most favourable is the nearest-neighbour site.
The intermediate site becomes less favourable in accord
with Lieb's theorem, which predicts a ferromagnetic solution of spin
2\,$\mu_{B}$.
The electronic structure of these paired chemisorbed hydrogen
atoms is such that the graphene-covered layer becomes semiconducting
for occupancy of nnNN sites at coverages near $\theta=2/9$
or greater, while it remains semimetallic for NN and nNN sites, even
for coverages up to $\theta=1/4$.
These results should help in the rational design
of hydrogen storage systems based upon carbon substrates:
a graphene layer exposed to molecular hydrogen would form
a physisorbed species with low barriers to diffusion. 
External compressive stress in the range between
$5$\% to $10$\% helps to dissociate molecules
to form atomic chemisorbed hydrogen. These are bound
to graphene with a low dissociation probability,
and diffuse slowly on the surface.
Releasing the external stress would induce the reverse
reaction, favouring the formation of molecular hydrogen,
with a low desorption barrier.

Financial support from the Spanish CYCIT is acknowledged
(MAT2008-1497, CSD2007-41 NANOSELECT, and FIS2009-8744).



\newpage

\quad \vfill

\vfill


\end{document}